\documentclass[twocolumn,pre,showpacs,amsfonts,amsmath,assymb,eufrak
,longbibliography]{revtex4-1}
\usepackage{epsfig}
\usepackage{color}
\usepackage{bm}
\usepackage[
colorlinks=true, 
pdfstartview=FitV, 
linkcolor=blue, 
citecolor=magenta, 
urlcolor=blue, 
bookmarks=true,
bookmarksnumbered=true,
pdftitle={},
pdfauthor={}
]{hyperref}
\usepackage{epsfig,amsopn}
\usepackage{graphicx}

\usepackage{amssymb}
\usepackage{amsmath}
\usepackage{color}
\usepackage{subfigure}
\usepackage{textcomp}
\usepackage[utf8]{inputenc}
\definecolor{pink}{rgb}{1,0,0.6}

\newcommand\ra{\rangle}
\newcommand\la{\langle}
\newcommand\nn{\nonumber}
\newcommand\f{\frac}
\newcommand\p{\partial}
\usepackage{grffile}
\graphicspath{{./Figures/}{./Hydro/}{./SpaceAverage/}}
\usepackage{textcomp}
\usepackage{amsmath}

\DeclareMathOperator\erf{erf}

\newcommand{\be}{\begin{equation}}
\newcommand{\ee}{\end{equation}}
\newcommand{\bea}{\begin{eqnarray}}
\newcommand{\eea}{\end{eqnarray}}

\begin{document}

\newcommand{\titlename}{Blast in a One-Dimensional Cold Gas: From Newtonian Dynamics to Hydrodynamics}

\title{\titlename}

\author{  Subhadip Chakraborti$^1$, Santhosh Ganapa$^1$, P. L. Krapivsky$^{2,3}$ and Abhishek Dhar$^1$}

\affiliation{$^1$International Centre for Theoretical Sciences, Tata Institute of Fundamental Research, Bengaluru 560089, India}%
\affiliation{$^2$Department of Physics, Boston University, Boston, Massachusetts 02215, USA}
\affiliation{$^3$Skolkovo Institute of Science and Technology, 143026 Moscow, Russia}

\date{\today}
\begin{abstract} 
A gas composed of a large number of atoms evolving according to Newtonian dynamics is often described by continuum hydrodynamics. Proving this rigorously is an outstanding open problem, and precise numerical demonstrations of the equivalence of the  hydrodynamic and microscopic descriptions are rare. We test this equivalence in the context of the  evolution of a blast wave, a problem that is expected to be at the limit where hydrodynamics could work. We study a one-dimensional  gas at rest with instantaneous localized release of energy for which the hydrodynamic Euler equations admit a self-similar scaling solution. Our microscopic model consists of hard point particles with alternating masses, which is a nonintegrable system with strong mixing dynamics. Our extensive microscopic simulations find a remarkable agreement with  Euler hydrodynamics, with deviations in a small core region that are understood as arising due to heat conduction.
\end{abstract}
\maketitle

The Navier-Stokes-Fourier (NSF) equations~\cite{zeytounian2012}  describing the evolution of the density, velocity, and temperature apply to an enormous range of phenomena, e.g., to atmospheric flows. At the fundamental level, however, molecules follow Newton's equations of motion. How accurate is the hydrodynamic description and what are the limits of its applicability? Here we address this question, which is of fundamental and practical importance. A rigorous derivation of the continuum hydrodynamics description from the atomistic one is an open problem~\cite{esposito2004,gallagher2019,gorban2018}  though there are phenomenological derivations via coarse-graining procedures~\cite{resibois1977,espanol2009,spohn2012}.  Much progress has been made using the Boltzmann equation  -- one derives the equations of hydrodynamics through a systematic expansion in a small parameter~\cite{chapman1990,de1989,saint2009}.  This still leaves one with the problem of deriving the Boltzmann equation from Newton's equations which has not been fully achieved even for dilute gases.  
For the case where one adds a weak noise to the Newtonian dynamics (still satisfying the same conservation laws) a rigorous derivation of the Euler equations for the hydrodynamic fields has been achieved \cite{olla1993}. 

Surprisingly,  there  appears to be no direct numerical verification that hydrodynamics accurately reproduces the predictions of the microscopic dynamics. The present work provides such a detailed comparison in the context of the classic blast-wave problem for which a self-similar scaling solution of the Euler equations was obtained more than sixty years back by Taylor~\cite{Taylor19501,Taylor19502}, von Neumann~\cite{VonNeumann1963} and Sedov~\cite{Sedov1946,Sedov2014}, and is referred to as the TvNS solution. 
The evolution of a blast wave emanating from an intense explosion was first studied to understand the mechanical effect of bombs. The rapid release of a large amount of energy in a localized region produces a surface of discontinuity beyond which the quantities concerned like density, velocity, and temperature fields change discontinuously~\cite{LandauBook,ZeldovichBook}.  The blast wave problem thus presents an extreme case to test the validity of hydrodynamics. 
From the point of microscopic models, the hard sphere system would be the natural candidate since much is known analytically. Second, this system can be simulated very efficiently using event-driven simulations. However, large scale molecular dynamics (MD) simulations \cite{Antal2008,Jabeen2010,Barbier2015,Barbier2015a,Barbier2016,Joy2017,Joy2019,Joy2021} have so far not found clear agreement with the TvNS  solution. It was suggested that  possible reasons for the differences could be the lack of local equilibration or due to the contribution of viscosity and heat conduction not included in the TvNS analysis. 

In this Letter, we address this question by studying hard point particles with binary mass distribution --- in one dimension, particles with equal masses just exchange velocities, so mass dispersion is necessary for relaxation, and the binary mass distribution is the simplest setting where relaxation is possible. Furthermore, we assume that adjacent particles have different masses (say $m_1$ and $m_2$). This alternating hard particle (AHP) gas has been extensively investigated in the context of the breakdown of Fourier's law of heat conduction in one dimension~\cite{Garrido2001,Dhar2001,Grassberger2002,Casati2003,Cipriani2005,Chen2014,Hurtado2016,Zhao2018,Lepri2020}.  The hard particle (and hard rod) system was investigated earlier in the context of the breakdown of the hydrodynamic description in one dimension~\cite{Kadanoff1995,Hurtado2006} and more recently the evolution of the AHP starting from a domain wall initial condition was studied in~\cite{Mendl2017} and incomplete thermalization of hard rods in a harmonic trap was observed in~\cite{moore2018}. Compared to the hard sphere system in higher dimensions, the 1D gas of point particles has several advantages -- the equation of state is  exactly that of an ideal gas, and simulations are faster since collisions occur only with nearest neighbors. Note that while the equilibrium physics is that of an ideal gas, the dynamics is nonintegrable and known to have good ergodic properties~\cite{casati1999}.

\begin{figure*}
	\begin{center}
		\leavevmode
		\includegraphics[width=5.8cm,angle=0]{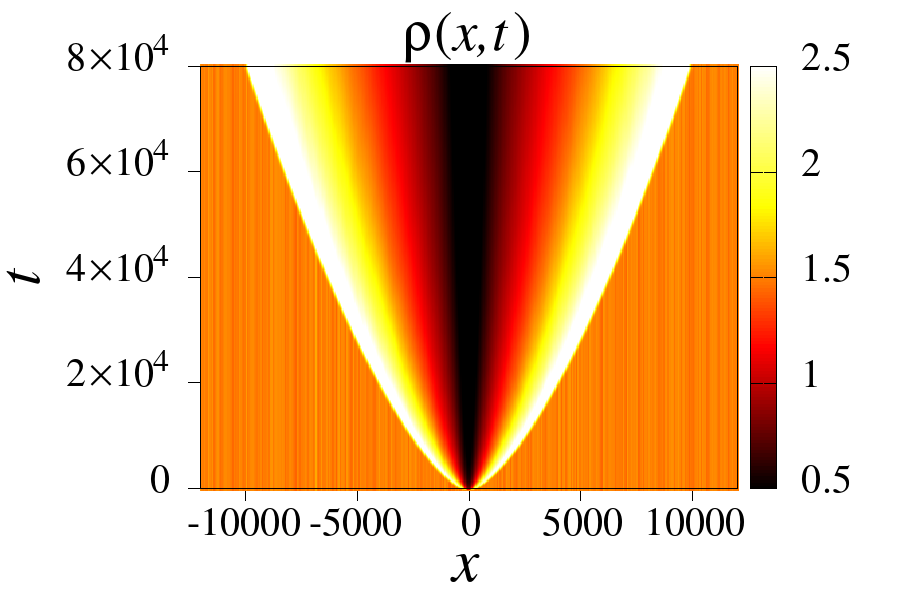}\hfill%
		\includegraphics[width=5.8cm,angle=0]{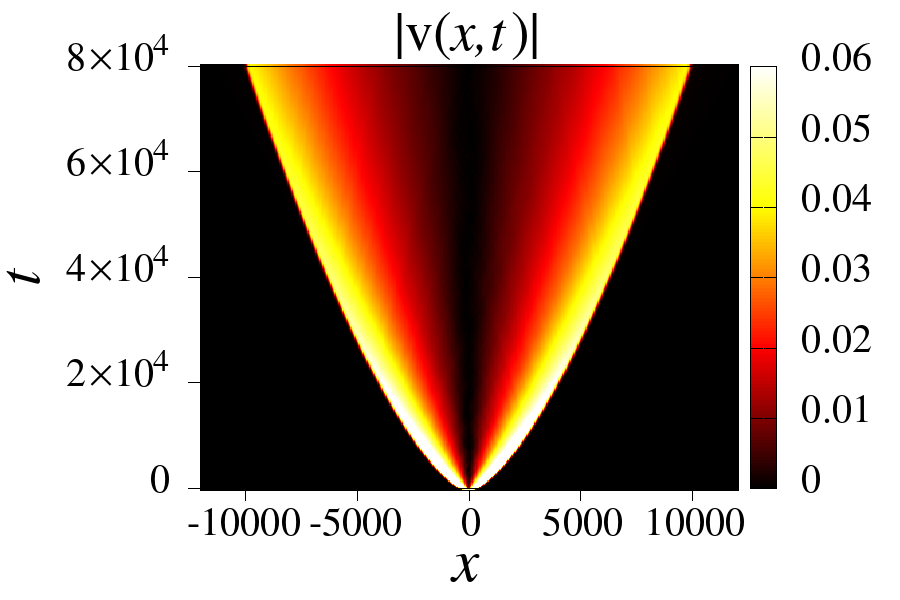}\hfill%
		\includegraphics[width=5.8cm,angle=0]{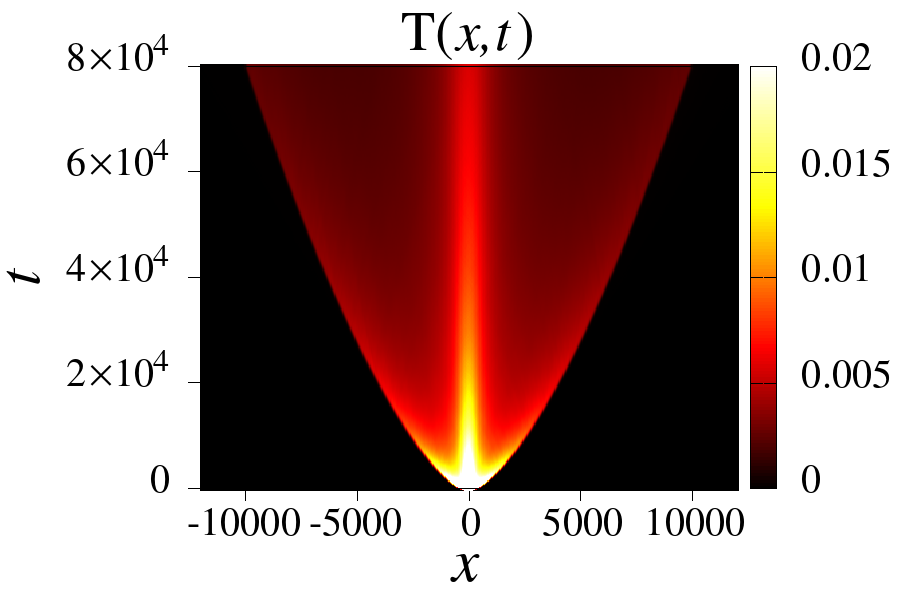}\hfill%
		\caption{Heat maps showing the spatiotemporal evolution of the  density, velocity and temperature fields, starting from initial conditions corresponding to a Gaussian initial temperature profile and $\rho(x,0)=\rho_\infty=1.5, v(x,0)=0$. The simulation parameters were $N=L=24000$, $E=32, \mu=1.5$, and an ensemble average was taken over $10^4$ initial conditions.}
		\label{heatmap}
	\end{center}
\end{figure*}

We study the evolution of the blast wave initial condition in the AHP gas. From extensive molecular dynamics simulations of the AHP gas, we compute the evolution of the density, velocity, and temperature fields and thereby extract the scaling forms obtained in the long time limit. We make comparisons with the TvNS scaling solution  which  we obtain exactly. We find that a complete explanation of the simulation of the blast requires us to go beyond the Euler equation and include the effect of heat conduction. We thus discuss the NSF equations, for which we present results from a numerical solution as well as a scaling analysis.

{\it The TvNS solution. --} We take a 1D ideal gas at zero temperature and uniform density $\rho_\infty$, and suddenly inject energy $E$  into a localized region of extent $\sigma$.    The  Euler equations for the density $\rho(x,t)$, velocity $v(x,t)$, and temperature $T(x,t)$ read
\begin{subequations}
\begin{align}
&\p_t \rho +\p_x (\rho v) =0,  \label{e11} \\
&\p_t (\rho v) +\p_x ( \rho v^2 + P ) =0, \label{e21} \\
&\p_t (\rho e) +\p_x (\rho v e + P v)  =0. \label{e31}
\end{align}
\end{subequations}
For an ideal gas $e=v^2/2+ T/(2\mu)$ and $P=\rho k_B T/\mu$, with $\mu=(m_1+m_2)/2$ for our binary mass system. We set Boltzmann's constant to unity: $k_B=1$.    A shock wave is formed, and it advances as $R \sim (Et^2/\rho_\infty)^{1/(d+2)}$ in $d$ dimensions \cite{LandauBook}, so in our 1D setting
\be
\label{shock_front}
R(t) =  \left(\f{E t^2}{A \rho_\infty}\right)^\frac{1}{3},
\ee
where $A$ is a dimensionless constant factor which, quite remarkably, can be computed exactly~\cite{supplemental}.  From dimensional analysis one can further show~\cite{LandauBook} that the fields will take the following  self-similar scaling form 
\begin{subequations}
\begin{align}
\rho(x,t)&=\rho_\infty G(\xi), \\
v(x,t)&=\f{2}{3}\f{x}{t} V(\xi)= \f{2\alpha}{3t^{1/3}} \xi V(\xi),\\
T(x,t)&=   \f{4 \mu}{27}\f{x^2}{t^2} Z(\xi)= \f{4\mu \alpha^2}{27 t^{2/3}} \xi^2 Z(\xi). \label{TvNS}
\end{align}
\end{subequations}
Here $\xi=x/R(t)$ is the scaled spatial coordinate, $\alpha = [E/(A \rho_\infty)]^{1/3}$  and the scaling functions $G,V,Z$ need to be determined. The factors $2/3,4/27$ are inserted for convenience; e.g., from Eq.~\eqref{shock_front} one finds that the velocity of the shock wave is $({2}/{3}) {R}/{t}$ and this suggests the use of the factor $2/3$.  Plugging these into Eqs.~(\ref{e11}--\ref{e31}) we find that the scaling functions satisfy a set of coupled first order ordinary differential equations in the variable $\xi$.  Using the condition of conservation of energy and the so-called Rankine-Hugoniot conditions, specifying the field discontinuities at the shock front,  allows one to obtain a complete closed-form solution of the problem, i.e., the functions $G,V,Z$ and the constant $A$~\cite{BarenblattBook}. In the Supplemental Material~\cite{supplemental} and Ref.~\cite{arxiv} we describe some details of the solution.

{\it Microscopic dynamics and initial conditions. -- }
Our system is a 1D  gas of $N$ hard point particles moving inside a box  $(-L/2, L/2)$. The only interactions between particles are through collisions between nearest neighbors that conserve  energy and momentum and also the ordering of the particles. Between collisions, the particles move ballistically with constant speeds while the postcollision velocities follow from Newton's laws~(see Ref. \cite{supplemental}).  For the AHP gas, the only conserved quantities are particle number, total momentum, and energy and we expect a hydrodynamic description in terms of the corresponding conserved fields that are obtained from the microscopic variables using the standard relations:
\begin{align}
(\rho,p,E)=\sum_{j=1}^N  \left\langle m_j (1,v_j,v_j^2/2) ~\delta[q_j(t) - x]\right\rangle,
\end{align}
where $\la ...\ra$ indicates an average over an initial distribution of microstates that correspond to the same initial macrostate. We define 
$v(x,t)=p/\rho$ and $e(x,t)=E/\rho$. For a nonintegrable system, it is expected that the evolving system is in local thermal equilibrium and the three fields $(\rho,v,e)$ should contain the local thermodynamic information at any space-time point $x,t$. Thus the internal energy per unit mass is $\epsilon(x,t)= e-{v^2}/{2}$, while ideal gas thermodynamics gives $T(x,t)= 2 \mu \epsilon(x,t)$ and $P(x,t)= 2 \rho(x,t) \epsilon(x,t)$.

We consider an initial macrostate where the gas has a  finite uniform density $\rho_\infty$, zero flow velocity $v$, and is at zero temperature everywhere except in a region of width $\sigma$ centered at $x=0$. This is the region of the blast and we take a smooth Gaussian profile  $E(x,0) = \frac{E}{\sqrt{2\pi\sigma^2}}e^{-x^2/{2\sigma^2}}$. The procedure to realize this macrostate in the microscopic simulations of the AHP gas and other details of our numerics are given in the Supplemental Material~\cite{supplemental}.

\begin{figure}
	\begin{center}
		\leavevmode
		\includegraphics[width=4.3cm,height=3.cm,angle=0]{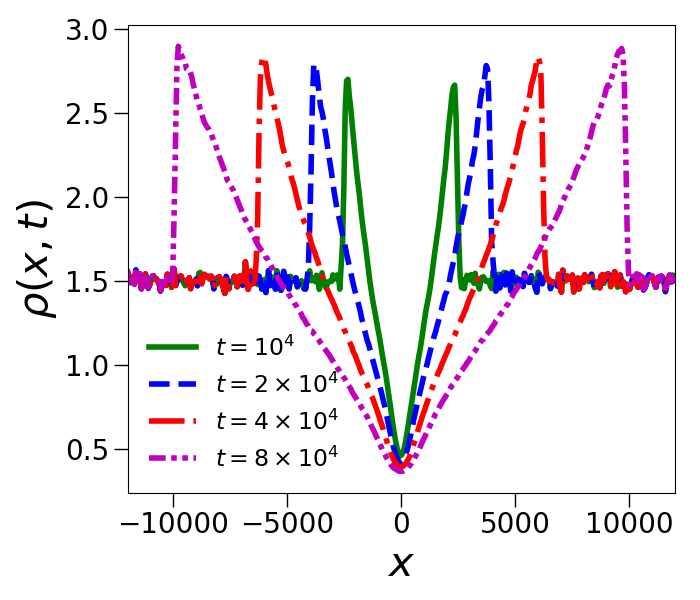}\hfill%
		\put (-125,75) {$\textbf{(a)}$}
		\includegraphics[width=4.3cm,height=3.cm,angle=0]{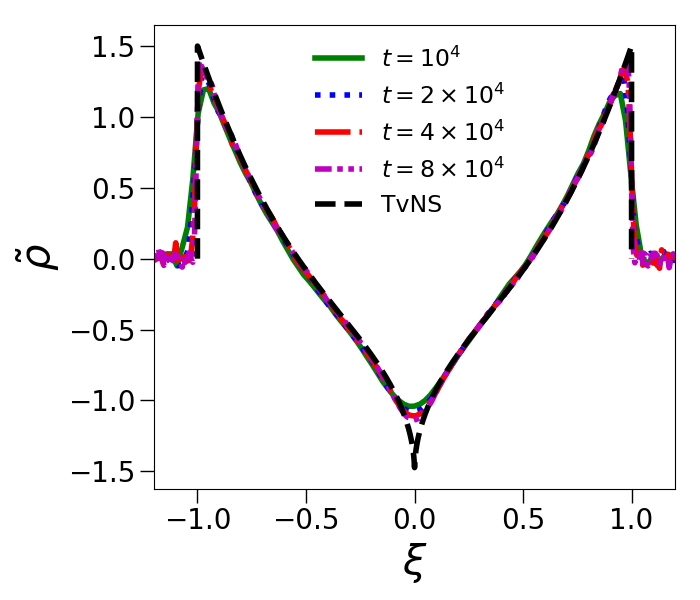}\hfill%
		\put (-120,75) {$\textbf{(d)}$}
				
		\includegraphics[width=4.3cm,height=3.cm,angle=0]{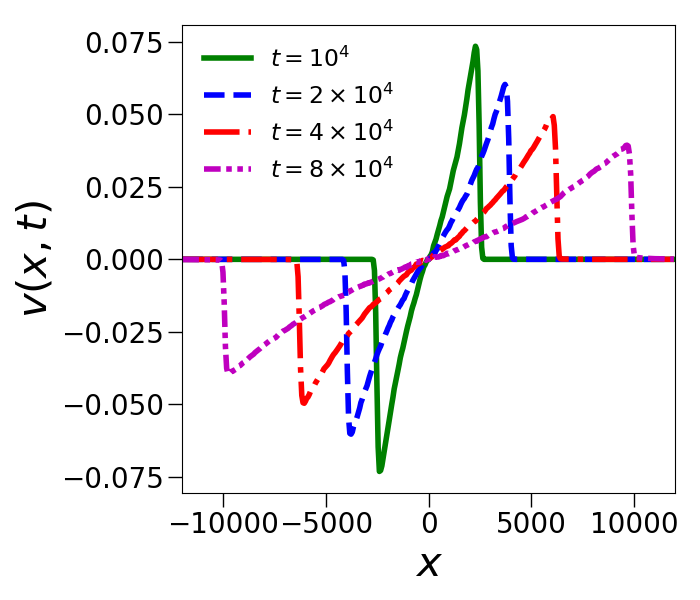}\hfill%
		\put (-125,75) {$\textbf{(b)}$}
		\includegraphics[width=4.3cm,height=3.cm,angle=0]{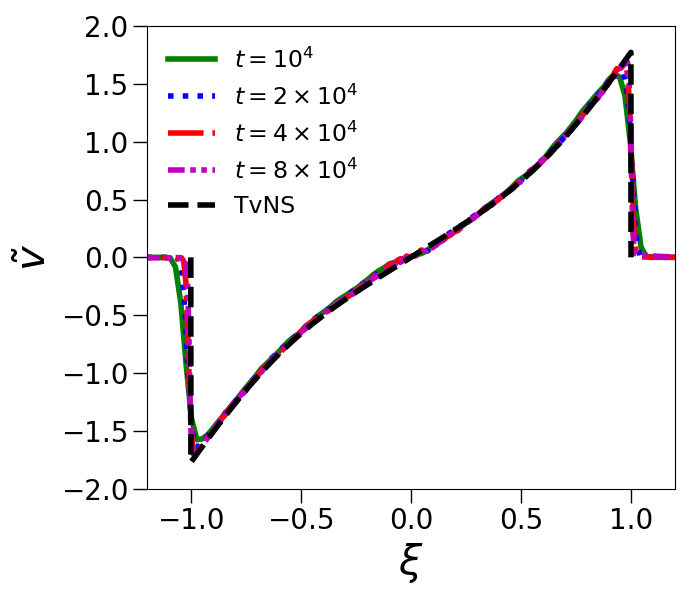}\hfill%
		\put (-120,75) {$\textbf{(e)}$}
				
		\includegraphics[width=4.3cm,height=3.cm,angle=0]{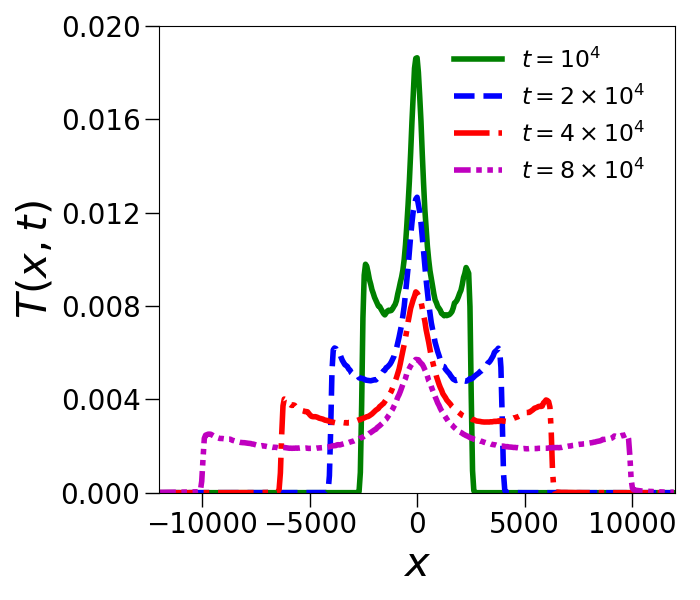}\hfill%
		\put (-125,75) {$\textbf{(c)}$}
		\includegraphics[width=4.3cm,height=3.cm,angle=0]{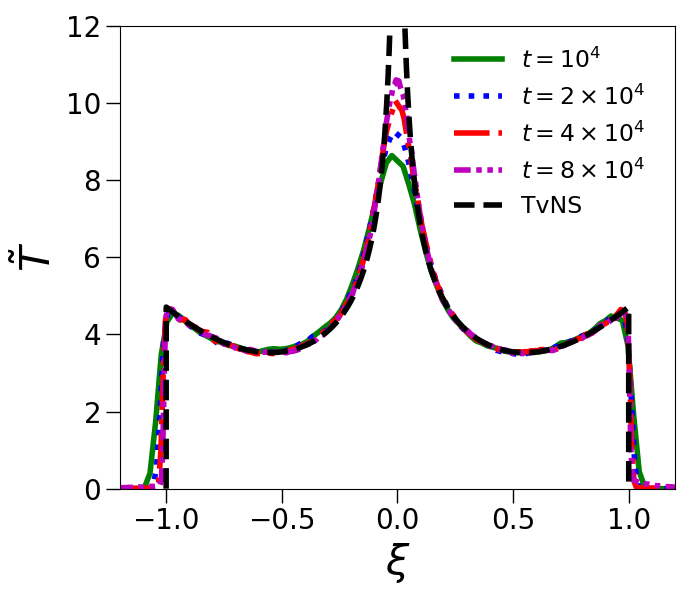}\hfill%
		\put (-122,75) {$\textbf{(f)}$}
		\caption{(a),(b),(c) Molecular dynamics results for the time evolution of density, velocity and temperature fields, starting from the initial conditions corresponding to a Gaussian initial temperature profile and $\rho(x,0)=\rho_\infty=1.5, v(x,0)=0$. The simulation parameters were $N =L=24000$, $E=32, \mu=1.5$ and an average was taken over $10^4$ initial conditions. (d),(e),(f) This shows the $x/t^{2/3}$ scaling of the data. We observe a very good collapse of the data at the longest times and an agreement with the TvNS scaling solution (dashed line).}
		\label{longtime}
	\end{center}
\end{figure}

\begin{figure}
	\begin{center}
		\leavevmode
		\includegraphics[width=4.3cm,height=3.cm,angle=0]{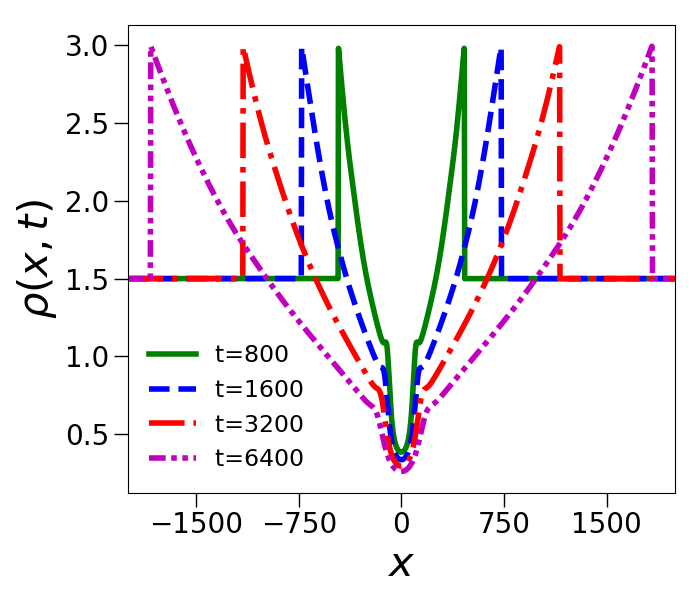}\hfill%
		\put (-125,75) {$\textbf{(a)}$}
		\includegraphics[width=4.3cm,height=3.cm,angle=0]{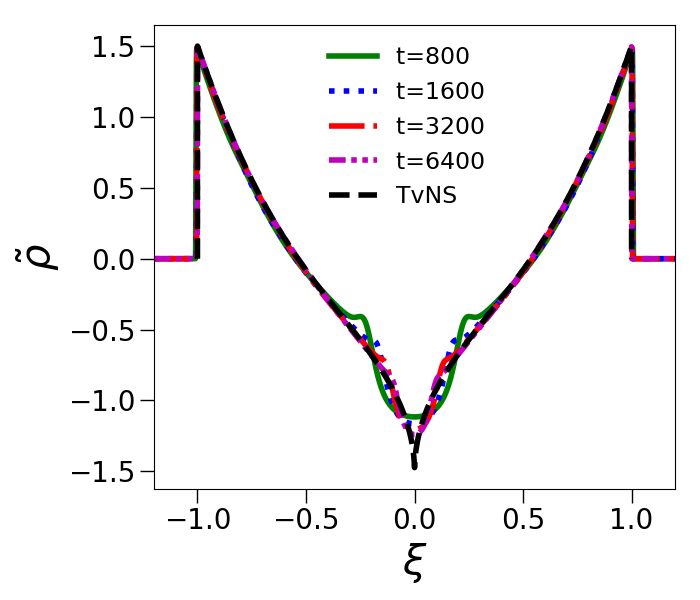}\hfill%
		\put (-120,75) {$\textbf{(d)}$}	
		
		\includegraphics[width=4.3cm,height=3.cm,angle=0]{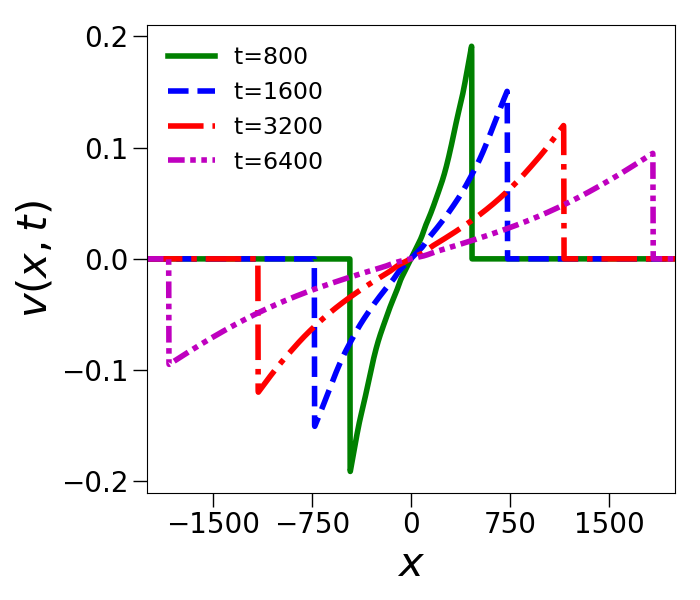}\hfill%
		\put (-125,75) {$\textbf{(b)}$}
		\includegraphics[width=4.3cm,height=3.cm,angle=0]{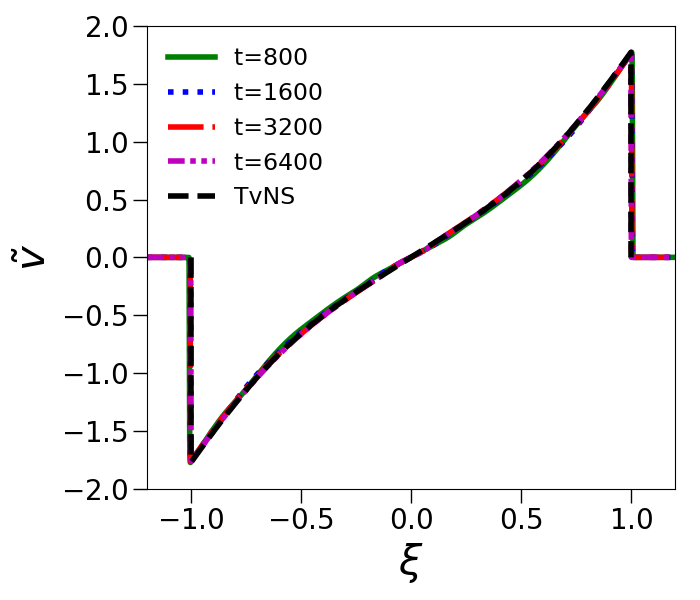}\hfill%
		\put (-120,75) {$\textbf{(e)}$}
			
		\includegraphics[width=4.3cm,height=3.cm,angle=0]{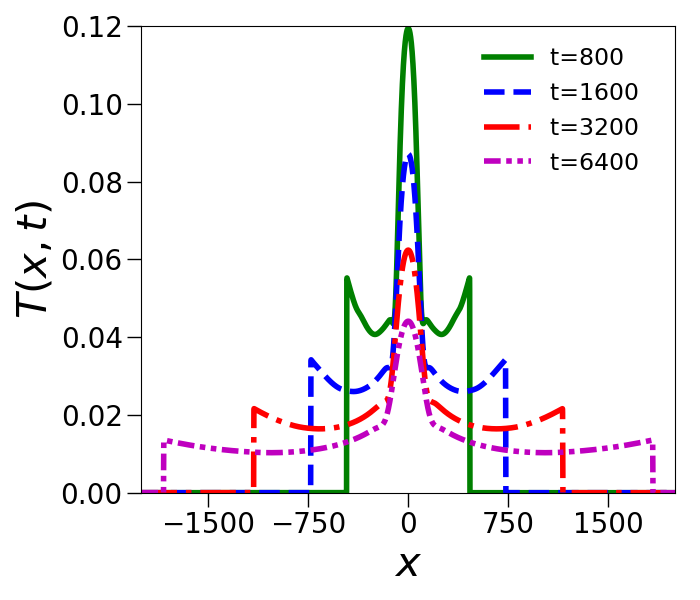}\hfill%
		\put (-125,75) {$\textbf{(c)}$}
		\includegraphics[width=4.3cm,height=3.cm,angle=0]{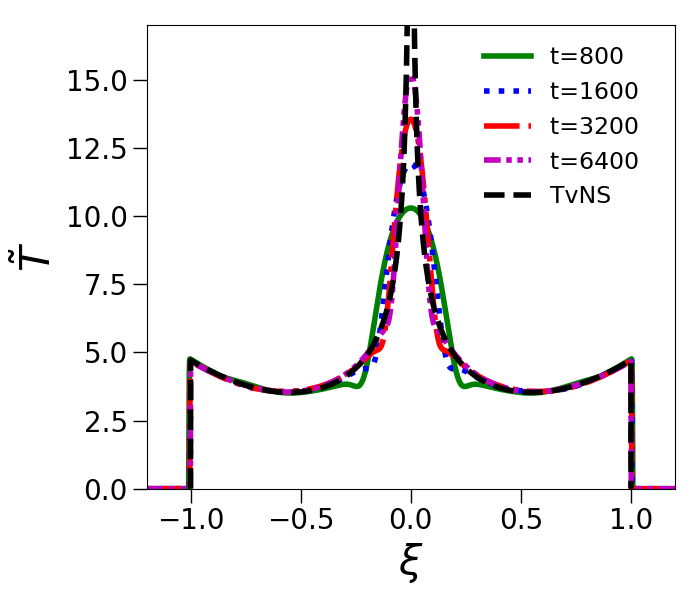}\hfill%
		\put (-122,75) {$\textbf{(f)}$}
		\caption{(a),(b),(c) Evolution of $\rho(x,t), v(x,t), T(x,t)$ obtained from a numerical solution of the NSF equations \eqref{tNS1}--\eqref{tNS3}, starting from the  same initial conditions as used  in the simulations for Fig.~\ref{longtime}. The other parameters used in the numerics  are $D_1=1$, $D_2=1$ and $L=4000$. (d),(e),(f) Scaling plot and comparison with the TvNS solution.}
		\label{noscaleHydnew}
	\end{center}
\end{figure}

{\it Comparison of simulations with the TvNS solution. --} 
We now present the results of the microscopic simulations for the evolution of the hydrodynamic fields  $\rho(x,t)$, $v(x,t)$, and $T(x,t)$ starting from the blast-wave initial conditions. In Fig.~\ref{heatmap} we show the spatiotemporal evolution of the  three fields (for individual particle trajectories see Ref.~\cite{supplemental}).  We see a sharp shock front that evolves sub-ballistically.  The mass density and flow velocity of the gas are peaked around the blast front while the temperature has an additional peak at the center.  In Figs.~\ref{longtime}(a),(b),(c) we plot the  evolution of the blast wave at long times.  In  Figs.~\ref{longtime}(d),(e),(f) we show the scaled fields  $\widetilde{\rho}=\rho_\infty G-\rho_\infty,\widetilde{v}=t^{1/3} v(x,t) =(2 \alpha/3) \xi V$ and $\widetilde{T}=t^{2/3} T(x,t)=(4 \mu \alpha^2/27) \xi^2 Z$ as  functions of the scaling variable $\xi$. We find an excellent collapse of the data everywhere except near the blast center. In the  region where there is a collapse of data, we find a  perfect agreement with the exact TvNS scaling solution (plotted as black dashed lines). Close to the origin, the TvNS solution predicts  the singular forms~\cite{arxiv} $G(\xi) \sim \xi^{1/2}$ and $Z(\xi) \sim \xi^{-5/2}$ which implies that near the origin the density vanishes as $\rho\sim x^{1/2}$ and the temperature diverges as $T\sim x^{-1/2}$. These are unphysical and disagree with simulations.  The deviations are caused by dissipation, specifically heat conduction that becomes important near the origin. Hence, we need to use the NSF equations.

{\it Comparison of simulations with numerical solution of the NSF equations. ---}
We  now  compare simulation results and the TvNS solution with those from the full dissipative hydrodynamic equations. In one dimension, the NSF equations read
\begin{subequations}
\label{NSF}
\begin{align}
&\partial_t \rho + \partial_x (\rho v) = 0,\label{tNS1} \\
& \rho (\partial_t  + v\p_x ) v + \partial_x (\rho T/\mu) = \p_x (\zeta\partial_x v), \label{tNS2} \\
& \f{\rho^3}{2 \mu} (\partial_t  + v\p_x ) \left(\f{T}{\rho^2}\right) = \p_x (\zeta v\partial_x v + \kappa\partial_x T ),
\label{tNS3}
\end{align}
\end{subequations}
where $\zeta$ denotes the bulk viscosity and $\kappa$ is the thermal conductivity of the system. These transport coefficients can depend on the fields and, based on the Green-Kubo relations, we expect their temperature dependence to be $\zeta \sim T^{1/2}$ and $\kappa \sim T^{1/2}$. A recent numerical study \cite{Hurtado2016} suggests the density dependence $\kappa \sim \rho^{1/3}$. In our numerical study we have thus used the forms $\zeta =D_1T^{1/2}$ and $\kappa=D_2\rho^{1/3}T^{1/2}$, where $D_1$ and $D_2$ are constants.

\begin{figure}
 	\begin{center}
 		\leavevmode
 		\includegraphics[width=4.2cm,height=3.2cm,angle=0]{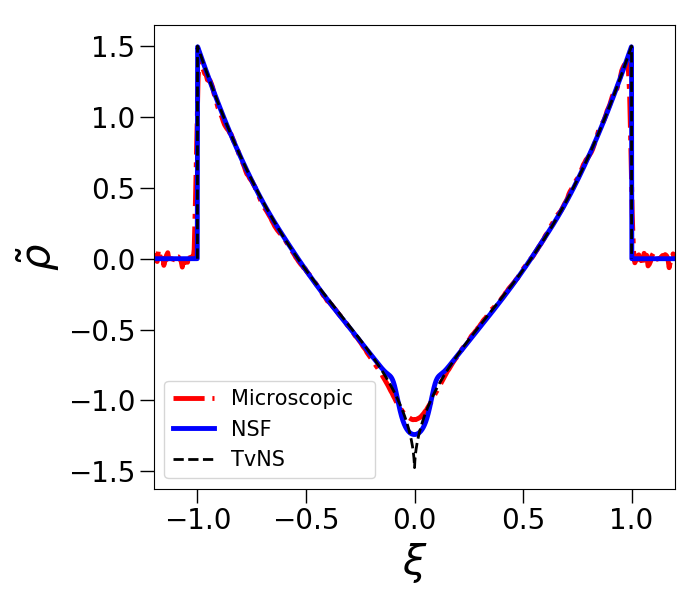}\hfill%
 		\put (-120,80) {$\textbf{(a)}$}
 		\includegraphics[width=4.2cm,height=3.2cm,angle=0]{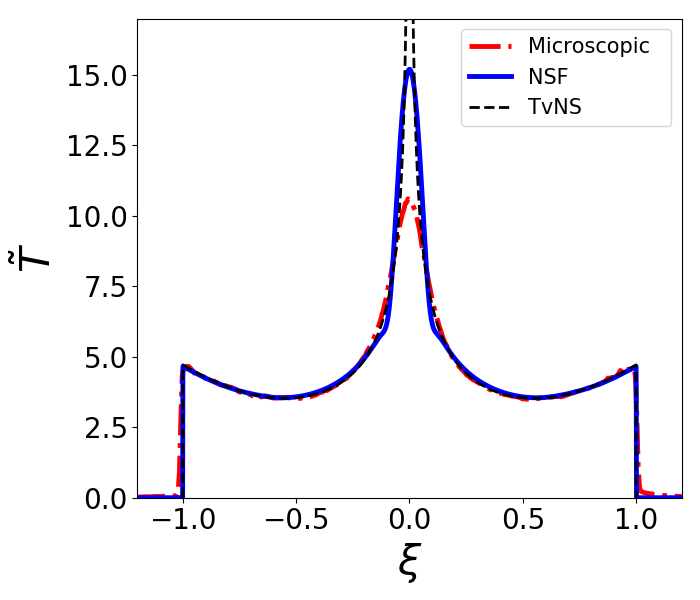}\hfill%
 		 \put (-120,80) {$\textbf{(b)}$}
 		\caption{Comparison of the scaled fields $\widetilde{\rho},\widetilde{T}$  from microscopic simulations (at time $t=80000$), 
		the NSF equations (at time $t=6400$), and the vNS scaling solution.}
 		\label{comparisonlong}
 	\end{center}
 \end{figure}

We solve Eqs.~\eqref{NSF} numerically~\cite{supplemental,maccormack1982} for the same initial conditions as considered in the microscopic simulations, namely $\rho(x,0)=\rho_\infty, v(x,0)=0$ and $E(x,0)$ given by the Gaussian form with total energy $E$. The numerical results are plotted in Figs.~\ref{noscaleHydnew}(a),(b),(c) which show $\rho(x,t)$, $v(x,t)$ and $T(x,t)$ at different times.  In Figs.~\ref{noscaleHydnew}(d),(e),(f) we plot the scaled fields $\widetilde{\rho},\widetilde{v},\widetilde{T}$ and verify the agreement with the TvNS solution everywhere except in the core. In the core, the NSF solution does not have any singularities, unlike the TvNS solution. In Fig.~\ref{comparisonlong} we plot together the long-time microscopic simulation results and the NSF results and it can be seen that in the core region, the NSF solution is in better qualitative agreement with simulations as compared to the TvNS solution.

We now estimate the size of the core and outline how to derive the corresponding scaling solution. Using Eq.~\eqref{tNS3} we see that the heat conduction term becomes important at a length scale $X$ such that $\rho\,\frac{T}{t} \sim \frac{T^{3/2}}{X^2}\, \rho^{1/3}$, where we neglect constant factors. 
The TvNS solution in the $\xi \to 0$ limit gives $\rho\sim \sqrt{\xi}\sim |x|^{1/2}/t^{1/3}$ and $T \sim \xi^{-1/2} \sim t^{1/3}/|x|^{1/2}$, from which we get an estimate
\begin{equation}
\label{size}
X\sim t^\frac{38}{93}
\end{equation}
for the size $X$ of the core where heat conduction is important. The outer solution (TvNS) and the inner (core) solution are comparable at $|x|=X$ and this allows us to determine the inner core scaling laws. Thus, the temperature at the center of the explosion is estimated from $T_0\sim X^{-\frac{1}{2}}\,t^{-\frac{1}{3}}$ and Eq.~\eqref{size}. Similarly, the density  at the center of the explosion can be estimated from $\rho_0\sim X^{1/2}/t^{1/3}$ and Eq.~\eqref{size}.  We thus arrive at 
\begin{equation}
\rho_0 \sim t^{-\frac{4}{31}}, \qquad  T_0 \sim t^{-\frac{50}{93}}
\end{equation}
while the velocity in the core scales as $X/t\sim t^{-{55}/{93}}$. 

\begin{figure}
	\begin{center}
		\leavevmode
		\includegraphics[width=4.2cm,height=3.2cm,angle=0]{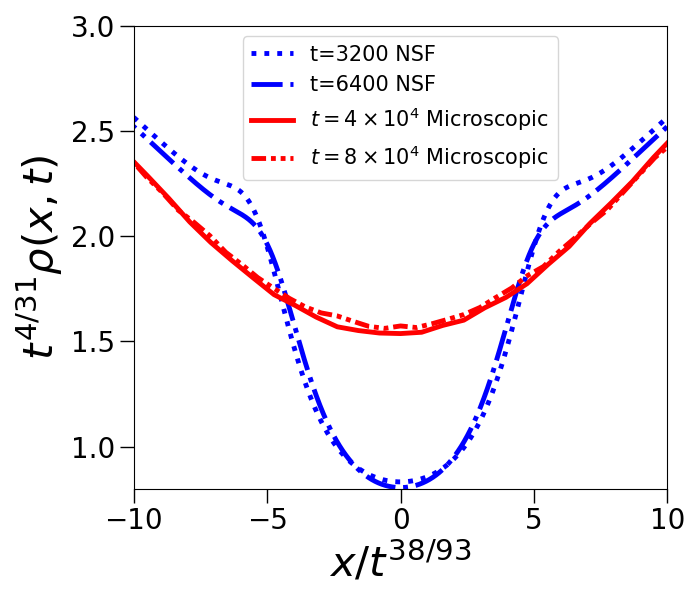}\hfill%
		 \put (-122,80) {$\textbf{(a)}$}
		\includegraphics[width=4.2cm,height=3.2cm,angle=0]{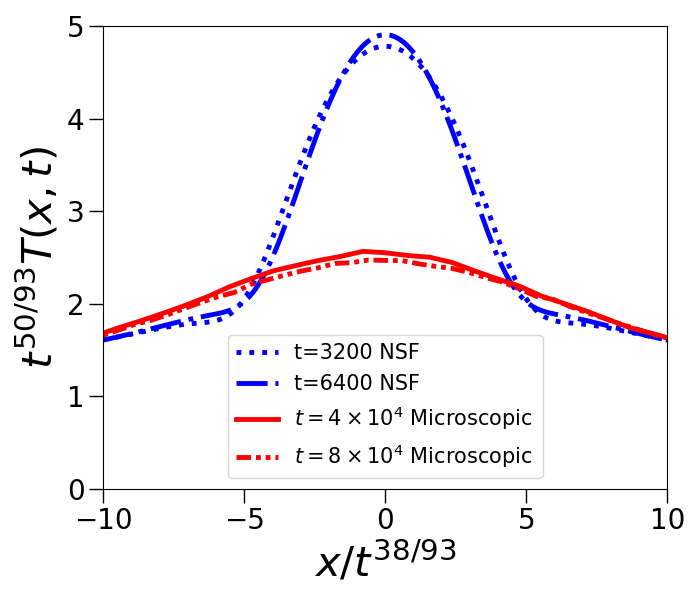}\hfill%
		 \put (-122,80) {$\textbf{(b)}$}
		\caption{Verification of the scaling form Eqs.~\eqref{scaling:T}, for $\rho$ and $T$, in the core of the blast. The comparison with data both from simulations and from the NSF solution are shown. The data  are the same as in  Figs.~\ref{longtime},\ref{noscaleHydnew}. The value $\eta=10 x/t^{38/93}$ corresponds to $\xi \approx 0.2$.}
		\label{fig:core}
	\end{center}
\end{figure}

The hydrodynamic fields in the hot core where dissipative effects play the dominant role should therefore exhibit scaling behaviors  in terms of the scaled spatial coordinate $\eta=x/X$. For large $x,t$, in a region with $\eta \sim \mathcal{O}(1)$, we expect the self-similar forms
\begin{equation}
\label{scaling:T}
 \rho = t^{-\frac{4}{31}} \widetilde{G}(\eta), \quad v = t^{-\frac{55}{93}} \widetilde{V}(\eta), \quad T = t^{-\frac{50}{93}} \widetilde{Z}(\eta).
\end{equation}
In Fig.~\ref{fig:core} we show that the data from the numerical solution of the NSF equations approximately satisfy this scaling form, though it appears that the convergence is somewhat slow. For comparison we also plot the simulation data at the last two times under the same scaling. A more detailed discussion of the inner-core scaling solution and its ``matching" with the outer-core solution can be found in Ref.~\cite{arxiv}.

{\it Conclusions. --} We have made a detailed comparison of the predictions of Euler hydrodynamics for a compressible one-dimensional gas with results from microscopic dynamics of the hard-core gas. Specifically, we have considered the blast problem, viz. a localized instantaneous release of energy in a cold gas.  
We have derived exact results for the front position and the hydrodynamic variables and found a remarkable agreement with microscopic simulations.  Deviations were seen in a core region whose size follows the scaling law with an unusual exponent, $X\sim t^{38/93}$. The position of the shock grows faster, $R\sim t^{2/3}$, so the relative size of the core region decays to zero in the long time limit. 
The width vanishes if measured in units of the TvNS scaling variable $\xi\sim x/t^{2/3}$. Thus we numerically establish that the Euler equations provide an asymptotically exact description of the hydrodynamic behavior of the AHP gas. This agreement is better than in higher dimensions \cite{Barbier2015,Barbier2015a,Barbier2016,Joy2017,Joy2019,Joy2021}. This is a bit counterintuitive as hydrodynamics is a mean-field theory that is expected to be more precise in higher dimensions. However, in one dimension we can consider a truly dilute system as the collisions are inevitable. Thus we know the exact equation of state in the AHP gas, and we obtained the TvNS scaling functions exactly. In higher dimensions, one relies on a virial expansion to get the equation of state and  the TvNS solution has to be numerically found.  The low-density ideal gas limit would require much longer simulation times. We have also checked that  unlike in the higher-dimensional cases~\cite{Joy2019,Joy2021} the local equilibrium, a key assumption in hydrodynamics, is accurately satisfied in our system~\cite{supplemental}.

The deviation at the core was understood as arising from the contribution of thermal conduction terms in the energy conservation equation, thus requiring a study of the full Navier-Stokes-Fourier equation for our 1D gas. For an accurate comparison with the simulation results, we needed the precise form of the thermal conductivity, $\kappa$, and in particular its dependence on temperature and density. For the hard particle gas, with the form $\kappa \sim \rho^{1/3} T^{1/2}$,  our analysis of the  NSF equations gives an estimate $|X| \sim t^{38/93}$ for the core size as well as scaling forms for the fields which we verified in the numerical solution of the NSF equations. The inclusion of the dissipative terms leads to a qualitative agreement between the microscopic simulation results and hydrodynamics even in the core region, in particular, it cures the erroneous prediction of the TvNS solution, the divergence of the temperature at the center of the explosion.  An open and challenging problem that remains is to obtain a quantitative agreement between simulations and hydrodynamics in the core region; this would require the precise form of the thermal conductivity of our 1D gas.  Although we focused on one dimension where we can provide a detailed numerical comparison, our approach can be extended to higher dimensions where we find a similar growing core region where thermal conductivity plays a role. The application of hydrodynamics to quantum systems is also of much recent interest~\cite{abanov2011,moore2020} and exploring this would be another interesting direction.

We thank Anupam Kundu, R. Rajesh, Sriram Ramaswamy, Samriddhi Sankar Ray and Vishal Vasan for useful discussions.
We acknowledge support of the Department of Atomic Energy, Government of India, under Project No. RTI4001.

\bibliographystyle{unsrt}
\bibliography{references}

\subsection*{Supplemental material for `Blast in a One-Dimensional Cold Gas: From Newtonian Dynamics to Hydrodynamics'}

\subsection{Exact TvNS scaling  solution of Euler equations}
\label{sec:TvNS}
We start with the equations
\begin{subequations}
\begin{align}
&\p_t \rho +\p_x (\rho v) =0,  \label{e1} \\
&\p_t (\rho v) +\p_x ( \rho v^2 + P ) =0, \label{e2} \\
&\p_t (\rho e) +\p_x (\rho v e + P v)  =0, \label{e3}
\end{align}
\end{subequations}
where we recall that $e=v^2/2+T/(2\mu)$ and $P=\rho T/\mu$. We seek a self-similar scaling solution of these equations with a shock at the location $R(t)$.  At the shock the condition of weak solutions lead to the  the following Rankine-Hugoniot boundary conditions
\begin{align}
&\f{\rho(R) v(R)}{\rho(R)-\rho_\infty }=U,  \\
&\f{ \rho(R) v^2(R) + P(R)}{\rho(R) v(R)}=U,  \\
&\f{ \rho(R) v(R) e(R) + P(R) v(R)}{\rho(R) e(R)}=U,
\end{align}
where $U=\dot{R}$. In addition, we need to use  the condition of energy conservation:
\begin{align}
\int_0^R dx \rho (v^2 + T/\mu) = E.    \label{ener}
\end{align} 
We seek the hydrodynamic variables in the scaling form
\begin{equation}
\label{scaling}
 \rho=\rho_\infty G(\xi),\quad v=\frac{2x}{3t}\,V(\xi),  \quad T=\f{4\mu }{27}\,\frac{x^2}{t^2}\,Z(\xi),
\end{equation}
where the re-scaled fields depend on the single dimensionless  variable
\begin{equation}
\label{xi-def}
\xi = \frac{x}{R}\,, \quad {\rm where}~R(t)= \left(\f{E t^2}{A \rho_\infty}\right)^{1/3}.
\end{equation}
Using the scaling form, the value $\gamma=3$ of the adiabatic index for the one-dimensional gas of point particles, and after some simplification, the Rankine-Hugoniot boundary conditions lead to
\begin{equation}
\label{BC}
G(1) = 2, \quad V(1) = \tfrac{1}{2}, \quad  Z(1) = \tfrac{3}{4}.
\end{equation}
Plugging the scaling forms \eqref{scaling}--\eqref{xi-def} into Eqs.~\eqref{e1}--\eqref{e2} and defining $\ell=\ln\xi$, we obtain 
\begin{align}
\label{VG-eq}
&\frac{dV}{d\ell}+(V-1)\,\frac{d\ln G}{d\ell}    = - V,  \\
\label{ZG-eq}
&\frac{d\ln Z}{d\ell} - 2\,\frac{d\ln G}{d\ell} = \frac{3-2V}{V-1}.
\end{align}
Using the fact that energy is conserved in the region $(-\xi R(t) < x < \xi R(t))$, one can express \cite{LandauBook,arxiv}  $Z$ through $V$ 
\begin{equation}
\label{ZV}
Z = \frac{3(1-V)V^2}{(3 V -1)}.
\end{equation}
The total energy conservation condition in Eq.~\eqref{ener} gives
\begin{eqnarray}
\label{energysc}
E  = \rho_\infty\, \f{4}{9} \,\frac{R^{3}}{t^2} \int_0^1 d\xi\,\xi^{2}\,\frac{2 V^3}{(3 V -1)}\,G,
\end{eqnarray}
where we have used \eqref{ZV}. Combining \eqref{energysc} with \eqref{xi-def} we obtain 
\begin{equation}
\label{A-int}
A =\f{4}{9}\, \int_0^1 d\xi\,\xi^{2}\,\frac{2V^3}{3 V -1}\,G.
\end{equation}
As shown in \cite{arxiv} it is possible to obtain an explicit solution of the set of equations~(\ref{VG-eq},\ref{ZG-eq},\ref{ZV}) with the boundary condition \eqref{BC}:
\begin{subequations}
\begin{align}
\label{xi-V:1}
&\xi^5 = 2^{-\frac{4}{3}}\,(3V-1)^2\, V^{-\frac{10}{3}}\,(3-4V)^{-\frac{11}{3}}\,,\\
\label{GV:1}
&G = 2^\frac{16}{5}\, (1-V)^2\, (3V-1)^\frac{1}{5}\, (3-4V)^{-\frac{11}{5}}.
\end{align}
\end{subequations}
Plugging this solution into Eq.~\eqref{A-int} we reduce the integral over $\xi$ to an integral over $V$ which is computed to give $A=152/1071$. Thus we have a completely closed form expression for the TvNS solution for the 1D ideal gas.

\subsection{Specification of microscopic dynamics and initial conditions}
\label{sec:IC}

 Let $q_j$, $p_j$, and $m_j$ be respectively the position, momentum, and mass of the $j^{th}$ particle with $j=1,2,\ldots,N$.  
For our binary mass model, we choose all even numbered particles to have mass $m_1$ and all odd ones $m_2$. Between collisions the particles move ballistically with constant speeds. After collision between particles $i$ and $i+1$ moving with velocities $v_i$ and  $v_{i+1}$, their post-collisional velocities follow from the conservation of energy and momentum: 
\begin{align}
v_i^\prime &= \frac{[2m_{i+1} v_{i+1} + v_i(m_i-m_{i+1})]}{(m_i+m_{i+1})}\,,\\  
v_{i+1}^\prime &= \frac{[2m_i v_i + v_{i+1}(m_{i+1}-m_i)]}{(m_i+m_{i+1})}\,.
\end{align}

In our numerical simulations we consider the following Gaussian profile for the initial energy:
\begin{align}
E(x,0) = \frac{E}{\sqrt{2\pi\sigma^2}}e^{-x^2/{2\sigma^2}}. 
~\label{eq:ezero}
\end{align} 
The total energy of the blast is $E$.

To obtain this in the particle model, we first distribute 
 $N$ particles labelled $i=-N/2+1,-N/2+2,\ldots,N/2$ uniformly between $x = -L/2$ to $L/2$,  so that the number density is $n_\infty=N/L=\rho_\infty/\mu$, where $\mu=(m_1+m_2)/2$ is the mean mass and $\rho_\infty$ the ambient mass density of the gas.  
For the $N_c$ particles around the center of the explosion with labels $i=-N_c/2+1, -N_c/2+2,\ldots N_c/2$, we choose their velocities from the Maxwell distribution with equal temperature: $\text{Prob}(v_i)= \sqrt{m_i/(2 \pi T_c)}  e^{-m_i v_i^2/(2 T_c)}$ with $T_c=2 E/N_c$. The velocities of other particles are set to zero. The size of the initial blast is thus approximately $s=N_c/n_\infty$.

We note that $E(x,0)= \la \sum_{i=1}^N  m_i v_i(0)^2/2 \delta(x-q_i(0)) \ra $.
 Since, for large $N$,  each particle's position is a Gaussian with mean $\bar{q}_i=i/n_\infty$ and variance $\sigma^2=L/(4 n_\infty)$, we then get:
\begin{align}
E(x,0) &= \sum_{i=-N_c/2+1}^{N_c/2} \f{T_i}{ 2 \sqrt{2 \pi \sigma^2}}  e^{-(x-\bar{q}_i)^2/ (2 \sigma^2)} \nn \\
&\approx \f{n_\infty T_i}{2} \int_{-s/2}^{s/2} dy   \f{e^{-(x-y)^2/ (2 \sigma^2)}}{  \sqrt{2 \pi \sigma^2}} \nn \\
&=\f{n_\infty T_i}{4} \left[ \erf{\left(\f{s-2x}{2 \sqrt{2} \sigma}\right)} +  \erf{\left(\f{s+2x}{2 \sqrt{2} \sigma}\right)}  \right]. \label{Eerf}
\end{align}
When $s \ll \sigma$ this  simplifies to 
\begin{align}
E(x,0)\approx E\, \f{e^{-x^2/ (2 \sigma^2)}}{  \sqrt{2 \pi \sigma^2}}. \label{Egauss}
\end{align}
We have verified that our long-time results are independent of the initial profile as long as it is localized and the total energy $E$ is fixed~\cite{arxiv}. 
\begin{figure}
	\begin{center}
		\leavevmode
		\includegraphics[width=8.cm,angle=0]{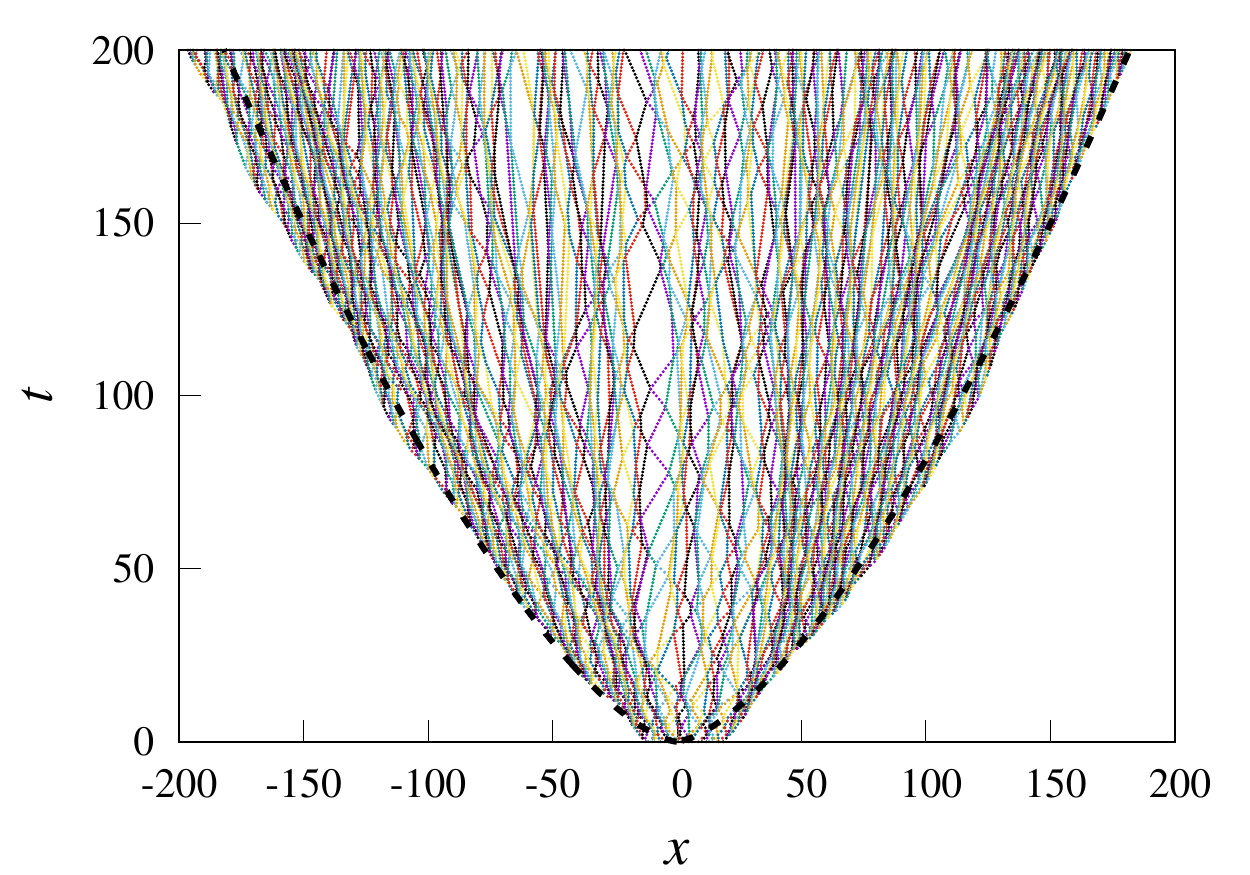}\hfill%
		\caption{{This plot shows the trajectories of $N=1000$ particles, starting from initial conditions corresponding to a Gaussian initial temperature profile and $\rho(x,0)=\rho_\infty=1.5, v(x,0)=0$. The expected shock front position $R(t)$ is shown as a dashed line. The stationary particles are not shown.}}
		\label{plot-trajectory}
	\end{center}
\end{figure}

The hydrodynamic fields  are obtained from the microscopic variables using the standard relations:
\begin{subequations}
\begin{align}
\rho(x,t)&=\sum_{j=1}^N  m_j \left\langle \delta(q_j(t) - x)\right\rangle, \\
p(x,t)&=\rho(x,t)v(x,t)=\sum_{j=1}^N  m_j \left\langle v_j \delta(q_j(t) - x)\right\rangle, \\
E(x,t)&=\rho(x,t)e(x,t)=\sum_{j=1}^N  \frac{m_j}{2}  \left\langle v_j^2 \delta(q_j(t) - x)\right\rangle.
\end{align}
\end{subequations}
Here $\la ...\ra$ indicates an average over an initial distribution of microstates that correspond to the same initial macrostate.

\subsection{Details of numerical techniques} 
\label{sec:numerics}
{\bf Molecular dynamics simulations:} In all our simulations we took $m_1=1, m_2=2$  (so that $\mu=(m_1+m_2)/2=1.5$), $\rho_\infty=1.5$ and $E=32, N_c=32$. In our  largest simulations we took $N=24000$, $L=24000$ and averaged over  an ensemble of ${\cal{R}}=10^4$ initial conditions.  For each microscopic initial condition, we evolved the system with the Hamiltonian dynamics. The molecular dynamics for this system was done using an event-driven algorithm that updates the system between successive collisions.

{\bf Solution of the NSF equations}: Our numerical solution relies on the MacCormack method~\cite{maccormack1982}, which is second order in both space and time. We have used a discretization $dx=0.1$ and $dt=0.001$. The diffusion constants in the dissipative terms were set to the values $D_1=D_2=1$ and the system size was taken as $L=4000$.  We have evolved the system up to time $t=6400$ which is before the energy reaches the boundary.  We have checked numerically the convergence of the method by using different values of $dx$ and $dt$. Note that for the hyperbolic Euler equations, with no dissipative terms, the numerical approach leads to strong dispersive effects near the shock front. The presence of $D_1$ and $D_2$ terms in the equations, in addition to being necessary to explain the  microscopic observations, also ensure that the solution does not develop dispersion and is stable at long times.

\begin{figure}
  \includegraphics[width=0.5\textwidth]{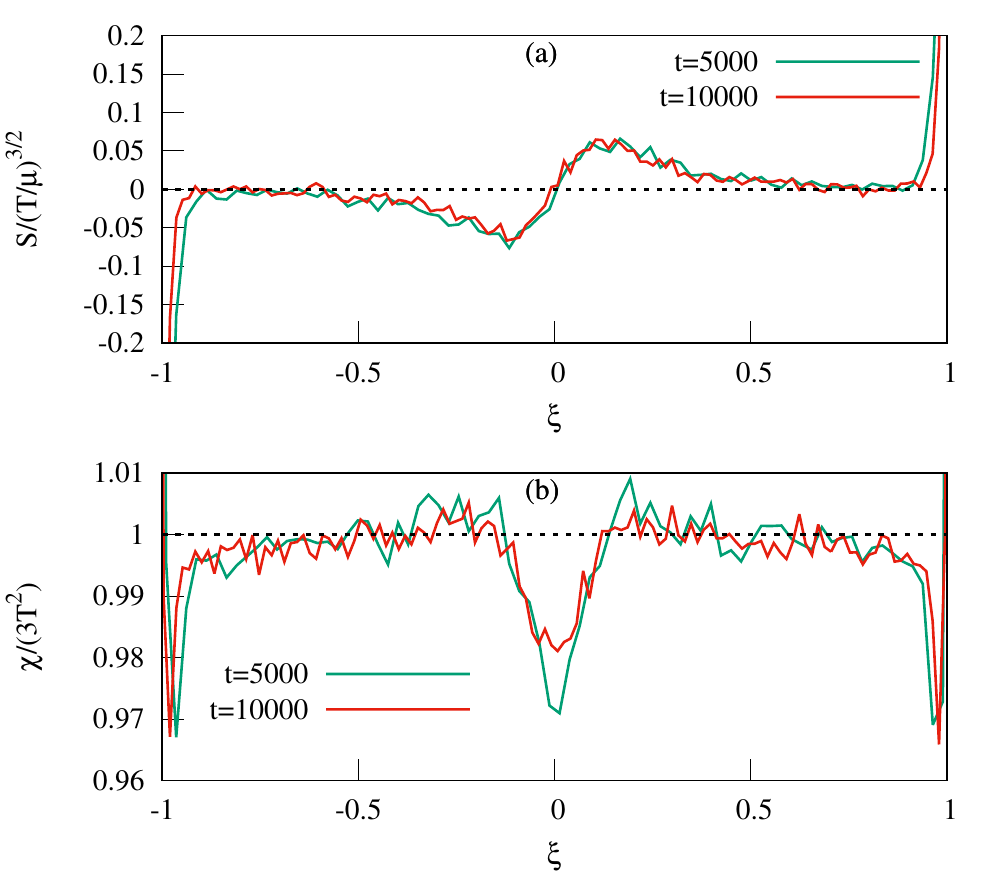}
  \caption{The spatial profile of the skewness defined via $S=n^{-1} \sum_i [v_i-u(x,t)]^3 \delta(x_i-x)$ and the kurtosis defined via $\chi= n^{-1} \sum_i m_i^2 [v_i-u(x,t)]^4 \delta(x_i-x)$ are plotted versus the scaled spatial coordinate $\xi=x/R(t)$. Here $n$ is the local particle number density. The skewness is scaled by $[{\rm thermal ~velocity}]^3$ while the curtosis is scaled by the expected thermal value of $3 T^2(x,t)$. The results at two different times are shown.}
\label{fig:le}
\end{figure}

\subsection{Probing the validity of the hydrodynamic description}

Hydrodynamics is expected to provide an adequate description for phenomena where the characteristic macro scale ($R$ in our case) greatly exceeds the micro scale (separations between adjacent particles in 1D). It is thus not expected to describe the structure of the shock wave since the width of the shock is its scale and the width is of the order of the separation (the micro scale). 

The local equilibrium (LE) assumption is an important requirement  for the validity of the hydrodynamic description and verifying this would be a direct probe of the validity of the hydrodynamic description.  Previous studies in two and three dimensions \cite{Joy2019,Joy2021} find strong deviations from the LE --- the observed skewness and kurtosis of the local velocity distributions indicate non-Gaussianity. The lack of the LE can be a reason for the observed disagreement with the hydrodynamic (TvNS) predictions.  Here we show results for these quantities obtained from MD simulations of our 1D system. Fig. \ref{fig:le} shows the LE assumption is satisfied quite accurately,  significantly better than in higher dimensions~\cite{Joy2019,Joy2021}. We do see small deviations from the LE which, interestingly, peak in the core and shock regions.  While this could provide a possible explanation for the disagreement between the NSF and simulations in the core region, we note however that we continue to have good agreement between these theories near the shock  front.

However, note that the structure of the shock wave is certainly not accounted by Euler hydrodynamics. Studying this and also the approach to the predictions of hydrodynamics is an avenue for future work. For example, finding the more precise formula for the position of the shock wave is an open problem. Hydrodynamics tells us that 
\begin{align}
R = [1071 E t^2/(152 \rho)]^{1/3}
\end{align}
but the deterministic hydrodynamic equations say nothing about the sub-leading term. The role of higher order terms in the Chapman-Enskog-Burnett gradient expansion~\cite{chapman1990}, towards determining corrections to the leading scaling form, also need to be probed further.

\end{document}